# Quantification of Human Movement for Assessment in Automated Exercise Coaching

Stuart Hagler, Holly B. Jimison, Ruzena Bajcsy, and Misha Pavel

*Abstract*— **Quantification of human movement is a challenge in many areas, ranging from physical therapy to robotics. We quantify of human movement for the purpose of providing automated exercise coaching in the home. We developed a model-based assessment and inference process that combines biomechanical constraints with movement assessment based on the Microsoft Kinect camera. To illustrate the approach, we quantify the performance of a simple squatting exercise using two model-based metrics that are related to strength and endurance, and provide an estimate of the strength and energy-expenditure of each exercise session. We look at data for 5 subjects, and show that for some subjects the metrics indicate a trend consistent with improved exercise performance.**

## I. INTRODUCTION

Physical exercise is important to the health and well-being of people of all ages. [1] The benefit of exercise to a person can be increased by a trained coach who is able to assess a person's ability, fatigue, and the difficulty of different moves. However, in practice, the ability of people to exercise with coaches is limited by the cost and availability of trained coaches and the demands of a person's personal schedule. We have, therefore, been working on the development of an automated coaching system to potentially provide a low-cost alternative to human coaches that can be available at that person's convenience and provide similar guidance. [2] The prototype system uses a Microsoft Kinect camera to make 3D measurements of a person's position over time during the performance of physical exercises, estimates the pose and the movements and attempts to provide appropriate feedback to the exercising person. A significant challenge in generating the most appropriate feedback is assessing the difficulty of the movements as well as the fatigue of the participant. The focus of this paper is to describe an approach to the development of model-based performance metrics related to these features.

In particular, in the present paper, we derive performance metrics related to strength and endurance for a single exercise called shallow squats, which is part of an exercise regimen for older adults that aims to improve mobility (i.e., getting into and out of a seated position and walking). [1] The exercise is simple and permits straight-forward analysis, but it can be generalized to a wider array of exercises. The data for our study came from exercise performed in the home as part of a coaching protocol, and the 3D motion data was collected using a Kinect camera.

## II. PERFORMANCE METRICS

### A. Shallow Squats Model

The exercise example analyzed in this paper is the shallow squats exercise, which is performed by assuming a comfortable standing posture with feet together. The person then lowers the torso by bending the knees and keeping the torso upright, lowering the hips as far as they can be comfortably lowered. Finally, the person returns the torso to the fully upright position. The shallow squat is illustrated in Fig. 1.

We model the motion in the shallow squats exercise using a single leg since both legs are typically constrained to make similar movements. For the sake of simplicity, in our

Manuscript received March 17, 2014, revised June 9, 2014.
This work was supported by the National Science Foundation (Grant 1111722).

Stuart Hagler is with the Department of Biomedical Engineering, Oregon Health & Science University, Portland, OR 97239 USA (e-mail: haglers@ohsu.edu).
Holly B. Jimison and Misha Pavel are with College of Computer & Information Science / Bouvé College of Health Sciences, Northeastern University, Boston, MA 02115 USA (e-mail: h.jimison@neu.edu; m.pavel@neu.edu).
Ruzena Bajcsy is with of the Department of Electrical Engineering & Computer Science at the University of California at Berkeley, Berkeley, CA 94720 USA (e-mail: bajcsy@eecs.berkeley.edu).

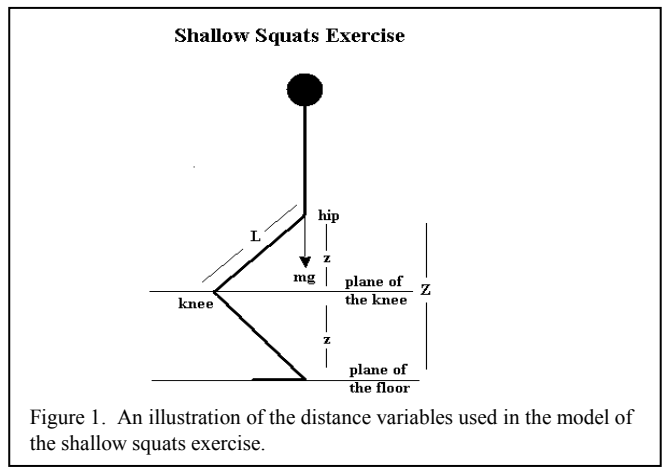

Figure 1. An illustration of the distance variables used in the model of the shallow squats exercise.

analysis, we neglect the feet and only consider the motion of the thigh which includes at one end the knee, and at the other end the hip which is approximated as a mass m at a point. The hip is constrained to move in the vertical axis above the heel. Due to the physical constrains on the motion, the position of the hip determines the position of the knee and thus the pose of the body at any time; so it is sufficient to only analyze motion relative to the plane of the knee which is defined to be the plane that is parallel to the floor and passes through the knee. Clearly, in the course of the exercise the plane of the knee moves up and down as the knee flexes, so to measure the motion relative to the floor we account for the movement of the plane of the knee. The torso is assumed to remain upright above the hip.

We indicate the position of the hip above the plane of the floor by real variable $Z$, the position of the hip above the plane of the knee by the real variable $z$, and the length of the thigh by the constant value $L$. Assuming that the length of the shin and the length of the thigh are the same, the position of the hip above the plane of the floor is related to the position of the hip above the plane of the knee by $Z = 2z$, and the position of the hip above the plane of the floor at the top of the squat is $Z_{top} = 2L$. We treat a squat as ending with the hip at a height $h$ above the plane of the knee, so the position of the hip above the plane of the floor at the bottom of the squat is $Z_{bottom} = 2h$. The orbit of a squat is given by a function of the position of hips above the plane of the knee as a function of time, $z(t)$; for a downward squats the orbit begins at a position $L$ and ends at a position $h$.

For a given height $z$ of the hip above the horizontal plane of the knee, the torque about the knee due to gravity is:

$$\tau = mg\sqrt{L^2 - z^2}. \tag{1}$$

In the case where $z$ is held at a constant value $z < L$, muscles must be engaged to compensate for the torque resisting the force of gravity. In the course of generating this torque, the muscles must perform metabolic work (i.e. consume fuel so as to generate the torque). We assume that, in general the amount of metabolic work done by the muscles increases as the required torque increases. While $z$ is held fixed, the total metabolic work done by the muscles increases with time. Mathematically, we treat this by associating a metabolic work rate with the torque by assuming (unknown) function $P(\tau)$. The metabolic work for fixed $z$ is now $W = P(\tau)T$. In general, if $z$ is varied over some time interval, the metabolic work associated with compensating for the torque due to gravity is given by $W = \int_0^T P\left(mg\sqrt{L^2 - z^2}\right)dt$. In the absence of an exact expression for P we represent it over the limited range of movement in the shallow squats exercise by a second order Taylor expansion as $P(\tau) \approx p_0 + p_1\tau + \frac{p_2}{2}\tau^2$. We use the approximate Taylor series expansion of the square root (Eq. 1) given by $\sqrt{x} \approx c_0 + c_1 x$:

$$W \approx \left(p_0 + c_0 p_1 mg + \left(c_1 p_1 mg + \frac{p_2 m^2 g^2}{2}\right)L^2\right)T \\ - \int_0^T \left(c_1 p_1 mg + \frac{p_2 m^2 g^2}{2}\right)z^2 dt. \tag{2}$$

An underlying assumption is that the older adult system user is choosing shallow squats that minimize the total metabolic work given in a fixed time interval $T$. Given the model, the squat that does this is one that remains at height $z = L$ for time $0 \leq t < T$ and moves instantly to height $z = h$ at time $T$ with the whole squat requiring zero metabolic work. This movement is not physically reliable. If we would like to have a squat model that has a minimum metabolic work squat with metabolic work greater than zero that also takes a more physical orbit, we need to include metabolic work terms in Eq. (2) that associate metabolic work rates with time derivatives of $z$. To handle this problem, we introduce a term in the integrand in the square of the jerk ($\dddot{z}$) that represents a cost associated with controlling the movement (see [3-6]); this gives a metabolic work of the form:

$$W \approx \left(p_0 + c_0 p_1 mg + \frac{1}{2}\left(2c_1 p_1 mg + p_2 m^2 g^2\right)L^2\right)T \\ + \int_0^T \left(\frac{1}{2}\mu\dddot{z}^2 - \frac{1}{2}\left(2c_1 p_1 mg + p_2 m^2 g^2\right)z^2\right)dt. \tag{3}$$

In principle, given the definitions that are to be given for the metrics, we now have enough information to calculate the metric values using the empirical orbit of the movement. However in practice, measurements with the Kinect have some amount of noise that would need to be accounted for in an integration of the empirical data. To minimize the effect of noise on the calculation of the metrics, we opted to calculate the metrics using a mathematical model of the orbit taken and a few key empirical measurements: $Z_{top}$, $Z_{bottom}$, and $T$.

To simplify the definition of the metrics without the need to estimate the parameters in Eq. (4), a complex optimization using calculus of variation, we assume that the subject makes a single squat beginning and ending at $z = L$ that takes on a sinusoidal orbit, or:

$$z = \frac{1}{2}(L + h) + \frac{1}{2}(L - h)\cos\left(\frac{2\pi}{T}t\right). \tag{4}$$

We further assume that during the time between squats, the subject does not expend any additional metabolic work beyond the background metabolic rate.

## B. Strength Metric

3D measurement of human movement only provides information about the kinematic aspects of the motion of the human body through space, without using any information about the forces needed to accomplish the movements.

For the sake of simplicity we assume that an individual's strength is given the peak force the individual can exert. We, therefore, define the strength metric of a movement to be the peak acceleration of the torso reached in the course of making the movement. The magnitude of the peak acceleration for the orbit in Eq. (4) is:

$$\frac{\ddot{z}_{max}}{2\pi^2} \approx \frac{L-h}{T^2}. \quad (5)$$

For a single downward or upward movement in a squat, we associate the strength metric $s$ with the peak acceleration given by:

$$s = \frac{\ddot{z}_{max}}{\pi^2} \approx \frac{Z_{top} - Z_{bottom}}{T^2}. \quad (6)$$

It should be evident that for a variety of models the peak acceleration of the movement is proportional to this value.

## C. Work Metric

Endurance is related to the work performed during the exercise, so before dealing with an endurance metric, we must estimate the work done during the exercise. As with the strength metric, we must produce a metric on the 3D motion data and with no information internal to the body.

We have already constructed a model of the metabolic work done during a squat, so it is just a matter of integrating that model over the movement given by Eq. (4). Ignoring the background metabolic work rate given by $p_0 + c_0 p_1 mg$, the metabolic work associated with an entire squat (i.e. squatting down and returning to the initial position) is:

$$W \approx 4\pi^6 \mu \frac{(L-h)^2}{T^5} + \frac{1}{2}(2c_1 p_1 mg + p_2 m^2 g^2) \times \left(L^2 - \frac{1}{4}(L+h)^2 - \frac{1}{8}(L-h)^2\right)T. \quad (7)$$

If the squats are done sufficiently slowly, the first term on the right-hand side of Eq. (7) will be negligible. In that case, the work associated with a single upward or downward squatting movement is half of this, so the work metric $w$ is:

$$w = 8(2c_1 p_1 mg + p_2 m^2 g^2)^{-1} W \\ \approx \left(Z_{top}^2 - \frac{1}{4}(Z_{top} + Z_{bottom})^2 - \frac{1}{8}(Z_{top} - Z_{bottom})^2\right)|T. \quad (8)$$

Unlike the strength metric, which is relatively independent of the underlying assumptions and models, the form of Eq. (8) depends on the assumed orbit of the squat in Eq. (4); so other models will produce somewhat differing metrics.

## D. Endurance Metric

Endurance relates to the total amount of metabolic work that the individual is able to do over the course of the exercise session; so we define the endurance metric $e$ to be the sum of the metabolic work metrics over a series of movements $e = \sum w$.

## III. EMPIRICAL STUDY

Five older adults (81 ± 6.5 years) participated in a study where an automated exercise coaching system was placed in their homes. The exercises to be performed were indicated to the subjects using an interactive video displayed on a computer screen. 3D movement measurements were made using a Kinect. The exercise regimen coached by the automated system consisted of twelve separate exercises including shallow squats. Subjects were able to do the full regimen in a given order, or choose individual exercises by navigating a computer menu.

Each exercise in the regimen was preceded by a video in which the exercise was described in detail to the subject by a trained (human) coach. In subsequent sessions, subjects were free to skip this video if they were already familiar with the exercise. When subjects were performing the exercise the screen showed a video of the coach performing the exercise in a format similar to those of commercial exercise videos.

3D measurements based on the Kinect skeleton representation were made and stored when the subject was performing the exercises. A session of each of the twelve exercises required the subject complete ten repetitions of each exercise. An algorithm for detecting each exercise repetition using the 3D measurements was implemented, and subjects were given feedback whenever they completed a repetition of the exercise that had been detected by the system. A session continued until the system detected ten repetitions or the subject chose to end the session. Subjects were free to continue the session beyond ten repetitions if they chose. Initial metrics of performance were implemented based on the algorithm used to detect repetitions of the exercises to provide feedback about changes in performance of the exercises across sessions that could be used to encourage subjects. None of the performance metrics derived here were used for feedback in this preliminary study.

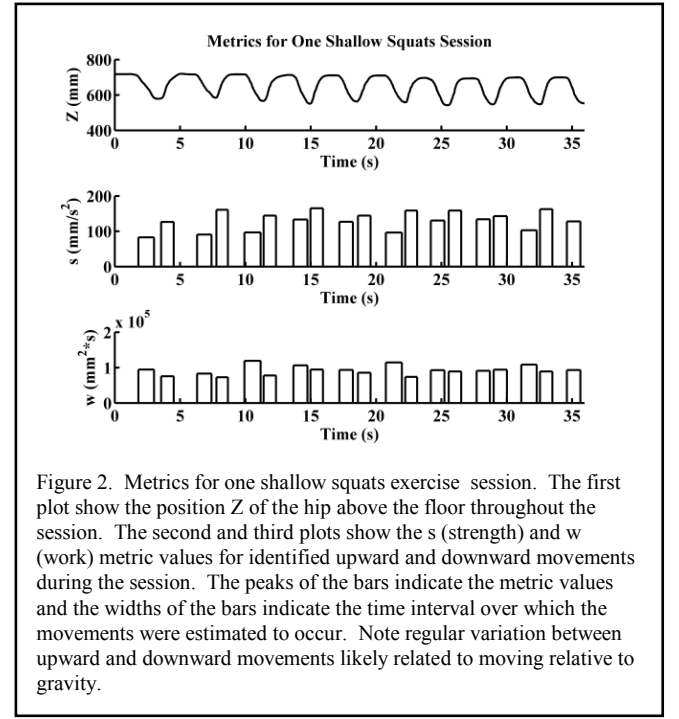

Figure 2. Metrics for one shallow squats exercise session. The first plot show the position Z of the hip above the floor throughout the session. The second and third plots show the s (strength) and w (work) metric values for identified upward and downward movements during the session. The peaks of the bars indicate the metric values and the widths of the bars indicate the time interval over which the movements were estimated to occur. Note regular variation between upward and downward movements likely related to moving relative to gravity.

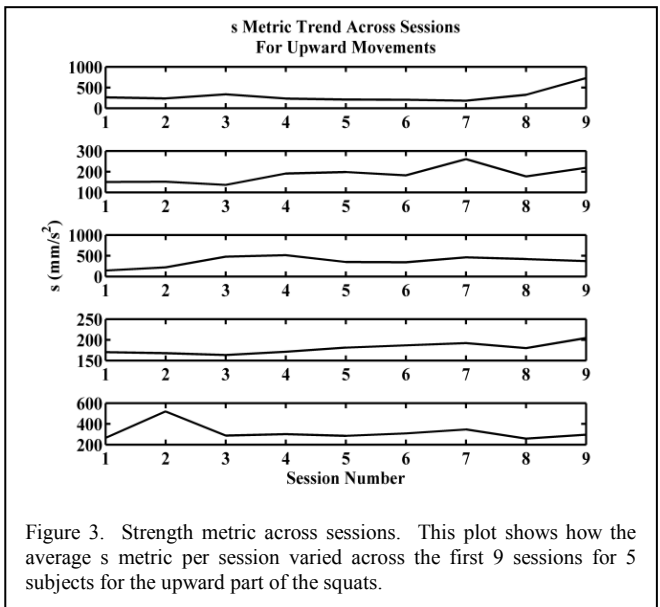

Figure 3. Strength metric across sessions. This plot shows how the average s metric per session varied across the first 9 sessions for 5 subjects for the upward part of the squats.

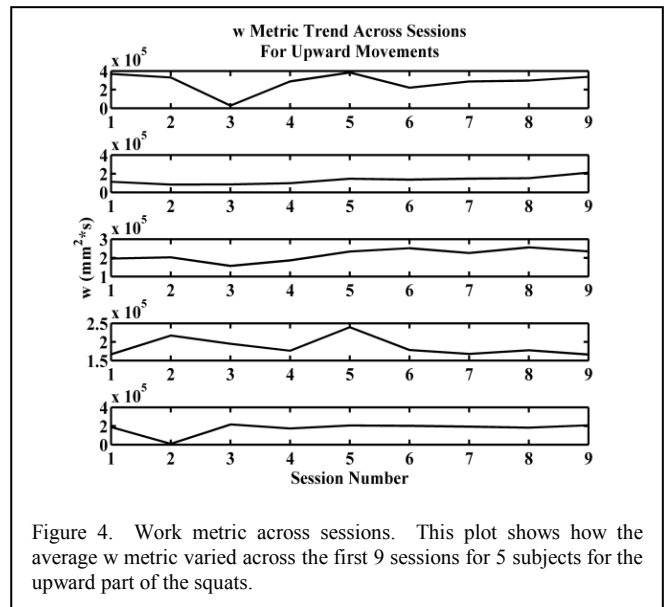

Figure 4. Work metric across sessions. This plot shows how the average w metric varied across the first 9 sessions for 5 subjects for the upward part of the squats.

For each session of the shallow squats exercise a simple algorithm was used to identify individual upward and downward movements within the sequence of ten squats in the session. We used only the recorded position of the hips. The data were smoothed, and by looking at the distribution of the vertical hip velocity through the session, a velocity threshold was identified. All intervals in which the vertical hip velocity exceeded the threshold were determined to be movement intervals, and the remaining intervals were determined to be pauses between movements. The values of the performance metrics were then calculated for each identified movement interval; this is illustrated by the data for a single session by a single subject in Fig. 2.

For the remainder of the analysis we restricted ourselves to data from upward movements to avoid any issues with the subject allowing themselves to fall under gravity during the downward portion of a squat.

To understand how the performance metrics indicated changes in exercise performance across sessions, we characterized each session by the average value of each of the s and w metrics. The results across the first 9 observed sessions are shown in Figs. 3 and 4; subjects are ordered in the same way from the top in both figures. We limited to analysis to 9 sessions to have the same amount of data for each subject (the subject that performed the fewest exercise sessions performed only 9). The s-metric appears to trend upward for second and fourth subjects from the top in Fig. 3 and is ambiguous or unchanging for the rest. The w-metric appears to trend upward for the second and third subjects from the top in Fig. 4 and is ambiguous or unchanging for the rest.

IV. CONCLUSION

We have presented two exercise performance metrics hypothesized to be related to strength and endurance for use with a Kinect-based automated exercise coaching system. These metrics appear to indicate trends in some subjects' performance of an exercise across a series of exercise sessions, and so potentially provide a means of quantifying changes in subject performance of the exercises over time which can be used to provide subjects with session-to-session-level feedback on exercise performance. Potentially, the quantification provided by the exercise metrics can be further refined to provide feedback at the repetition-to-repetition-level of an exercise, bringing the feedback closer to the feedback human coach would provide when observing the subject exercising in person.